\documentclass[seceq]{ptptex}

\usepackage{graphicx}

\usepackage{amsmath}

\newcommand{\ima}{{\mbox{Im}\,}}
\newcommand{\rea}{{\mbox{Re}\,}}

\title{Properties of light resonances from unitarized Chiral perturbation theory: Nc behavior and quark mass dependence
}

\author{
J. R. \textsc{Pel\'aez}$^a$,\footnote{speaker. J.R.P. thanks the NFQCD2010 organizers for the invitation and
for their work to create such an exciting workshop.}
J. \textsc{Nebreda}$^a$,
G. \textsc{R\'ios}$^a$
}

\inst{
$^a$Dept. de F\'{\i}sica Te\'orica II. Universidad Complutense. 28040 Madrid. Spain
}

\abst{
We review the unitarization of
Chiral Perturbation Theory with dispersion relations and how
it describes  meson-meson scattering data, generating
light resonances whose mass, width and nature can be related to QCD
parameters like quark masses and the number of colors.
}

\begin{document}

\maketitle
\vspace*{-.2cm}
\section{Introduction}
\vspace*{-.2cm}

Light hadron spectroscopy lies beyond the applicability of
perturbative QCD. However, there is an effective field theory, 
known as Chiral Perturbation Theory\cite{chpt1} (ChPT), which provides a description of
the dynamics of the lightest mesons. 
Despite it is limited to low energies and masses,
here we review how, when combined with
dispersion relations,  it leads to a successful description of meson dynamics, 
generating resonant states without a priori assumptions on their existence or nature.
This ``unitarized ChPT'' is a useful tool to identify
the spectroscopic nature of resonances through
their dependence on the QCD number of colors $N_c$, 
but also to relate lattice results to physical 
resonances by studying their quark mass, $m_q$, dependence.

ChPT is built out of the Goldstone Bosons of the QCD 
spontaneous chiral symmetry breaking, namely, pions, kaons and etas, 
as a low energy expansion 
of a Lagrangian respecting all QCD symmetries.
It is organized in powers of
 $p^2/\Lambda^2$, where $p$ stands either for derivatives, momenta or meson masses,
and $\Lambda\equiv 4 \pi f_\pi$, where
$f_\pi$ denotes the pion decay constant.
ChPT is renormalized order by order by absorbing loop
divergences in the renormalization of parameters of higher order
counterterms, known as low energy constants (LECs)
 that \emph{carry no energy or mass dependence} and depend on a 
regularization scale $\mu$. As always after renormalization,
the full amplitude is independent of this scale. 
Their  values depend
on the QCD dynamics, and are determined from
experiment.
Up to the desired order,  the ChPT expansion
provides a {\it systematic and model independent} 
description of how meson observables depend on QCD
parameters like the light quark masses $\hat m=(m_u+m_d)/2$ and $m_s$,
or the leading $1/N_c$ behavior \cite{'tHooft:1973jz}.


\vspace*{-.2cm}
\section{Dispersion relations and unitarization}
\vspace*{-.2cm}

Elastic resonances appear as poles on the
second Riemann sheet of the meson-meson
scattering partial waves $t_{IJ}$ of definite isospin $I$ 
and angular momentum $J$. 
At physical values of $s$, elastic unitarity implies 
\vspace*{-.2cm}
\begin{equation}
  \ima t_{IJ}(s)=\sigma (s) \vert t_{IJ}(s)\vert^2 
  \;\;\Rightarrow 
  \quad t_{IJ}=\frac{1}{\rea t_{IJ}^{-1} - i \sigma},
\quad {\rm with}
  \quad \sigma(s)=2p/\sqrt{s},
  \label{unit}
\end{equation}
where $s$ is the Mandelstam variable and $p$ is the
center of mass momentum. However, 
ChPT amplitudes, being an expansion
$t_{IJ}\simeq t_{IJ}^{(2)}+t_{IJ}^{(4)}+\cdots$, with
$t^{(2k)}=O(p^{2k})$, can only satisfy
Eq. (\ref{unit}) perturbatively
\begin{equation}
  \label{unitpertu}
  \ima t_{IJ}^{(2)}(s)=0,\;
  \ima t_{IJ}^{(4)}(s)=\sigma(s)\vert t_{IJ}(s)^{(2)}\vert^2,\;\dots
  \Rightarrow\;
  \ima t_{IJ}^{(4)}(s)/t_{IJ}^{(2)\,2}(s)=\sigma(s),
\end{equation}
and cannot generate poles. However, the resonance region can be
reached combining ChPT with dispersion theory either for the
amplitude~\cite{gilberto}
or for the inverse amplitude through the
Inverse Amplitude Method (IAM)~\cite{Truong:1988zp,Dobado:1992ha,GomezNicola:2007qj}. 
We will concentrate on the {\it one-channel} IAM 
\cite{Truong:1988zp,Dobado:1992ha}, since it uses ChPT 
only up to a given order inside 
a dispersion relation, without additional input or further model dependent assumptions. Other unitarization techniques will be commented below.

\vspace*{-.1cm}
\subsection{The one-loop ChPT Inverse Amplitude Method}
\vspace*{-.1cm}

For a partial wave $t_{IJ}(s)$, we can write a
dispersion relation (that we subtract three times,
since we will also use it below for $t_{IJ}^{(4)}$, that grows with $s^2$)
\begin{equation}
t_{IJ}(s)=C_0+C_1s+C_2s^2+
\frac{s^3}\pi\int_{s_{th}}^{\infty}\frac{\ima
t_{IJ}(s')ds'}{s'^3(s'-s-i\epsilon)} + LC(t_{IJ}).
\label{disp}
\end{equation}Note we have explicitly
written the integral over the physical cut, 
extending from threshold, $s_{th}$, to infinity, 
but we have abbreviated by $LC$ the equivalent
expression for the left cut (from 0 to $-\infty$). We could do similarly with 
other cuts, if present, as for $\pi K\rightarrow\pi K$.
Note that from Eq.\eqref{unit}
the imaginary part of the {\it inverse amplitude} is {\it exactly} known in the elastic regime.
We can then write a dispersion relation like that in \eqref{disp} 
but now for the auxiliary function $G=(t_{IJ}^{(2)})^2/t_{IJ}$, i.e.,
\begin{equation}
G(s)=G_0+G_1s+G_2s^2+     \\   \nonumber
\frac{s^3}\pi\int_{s_{th}}^{\infty}
\frac{\ima G(s')ds'}{s'^3(s'-s-i\epsilon)}+LC(G)+PC,
\label{Gdisp}
\end{equation}
where now $PC$ stands for possible pole contributions in $G$ coming from 
zeros in $t_{IJ}$. It is now straightforward to expand the subtraction constants
and use that
 $\ima t_{IJ}^{(2)}=0$ and  $\ima t_{IJ}^{(4)}=\sigma\vert t_{IJ}^{(2)}\vert^2$,
so that $\ima G= -\ima t_{IJ}^{(4)}$. In addition,
 up to the given order, $LC(G)\simeq -LC(t_{IJ}^{(4)})$,
whereas $PC$ is of higher order and can be neglected. Then
\begin{eqnarray}
\frac{t_{IJ}^{(2)2}}{t_{IJ}}\simeq a_0+a_1s-b_0-b_1s-b_2s^2 
-\frac{s^3}\pi\int_{s_{th}}^{\infty}\frac{\ima
t_{IJ}^{(4)}(s')ds'}{s'^3(s'-s-i\epsilon)}-LC(t_{IJ}^{(4)})
\simeq t_{IJ}^{(2)}-t_{IJ}^{(4)}, \nonumber
\label{preIAM}
\end{eqnarray}
since the $a_i, b_i$ terms, coming from
the $G_i$ expansion, are the subtraction terms of a dispersion relation for
$t_{IJ}^{(2)}-t_{IJ}^{(4)}$.
Thus we arrive at the so-called IAM:
\begin{equation}
t_{IJ}\simeq
t_{IJ}^{(2)2}/(
t_{IJ}^{(2)}-t_{IJ}^{(4)} ),
\label{IAM}
\end{equation}
that provides an elastic amplitude satisfying unitarity and has the correct
ChPT expansion up to the order we have used. 
The $PC$ contribution has been calculated  explicitly \cite{GomezNicola:2007qj} and is not just formally suppressed, but numerically negligible
except near the Adler zeros, away from the physical region.
It is straightforward
to extend the IAM to other elastic channels or higher orders \cite{Dobado:1992ha}.
Naively, the IAM 
looks like replacing $\rea t_{IJ}^{-1}$ by its $O(p^4)$ 
ChPT expansion in \eqref{unit}, but 
 \eqref{unit} is only valid in the real axis, whereas our derivation allows
us to consider the amplitude in the complex plane and look for poles associated to  resonances.
Let us remark that, since ChPT is used \emph{only at low energies} 
in the dispersion relation, the IAM formula is
justified only up to energies where inelasticities become
important, even though ChPT does not converge at those
energies.
Only when the energy is close to the Adler zero one should use
a slightly modified version of the IAM \cite{GomezNicola:2007qj}.
Re-expanding the IAM,  ChPT is recovered up to the order it was used
as input, as well as partial contributions to higher order, but not the complete series--- see 
Ref.\citen{Gasser:1990bv} for
a discussion of this issue.

\begin{figure}
\begin{tabular*}{0.5\textwidth}{@{\extracolsep{-0.4cm}}cc}
  \includegraphics[scale=.48]{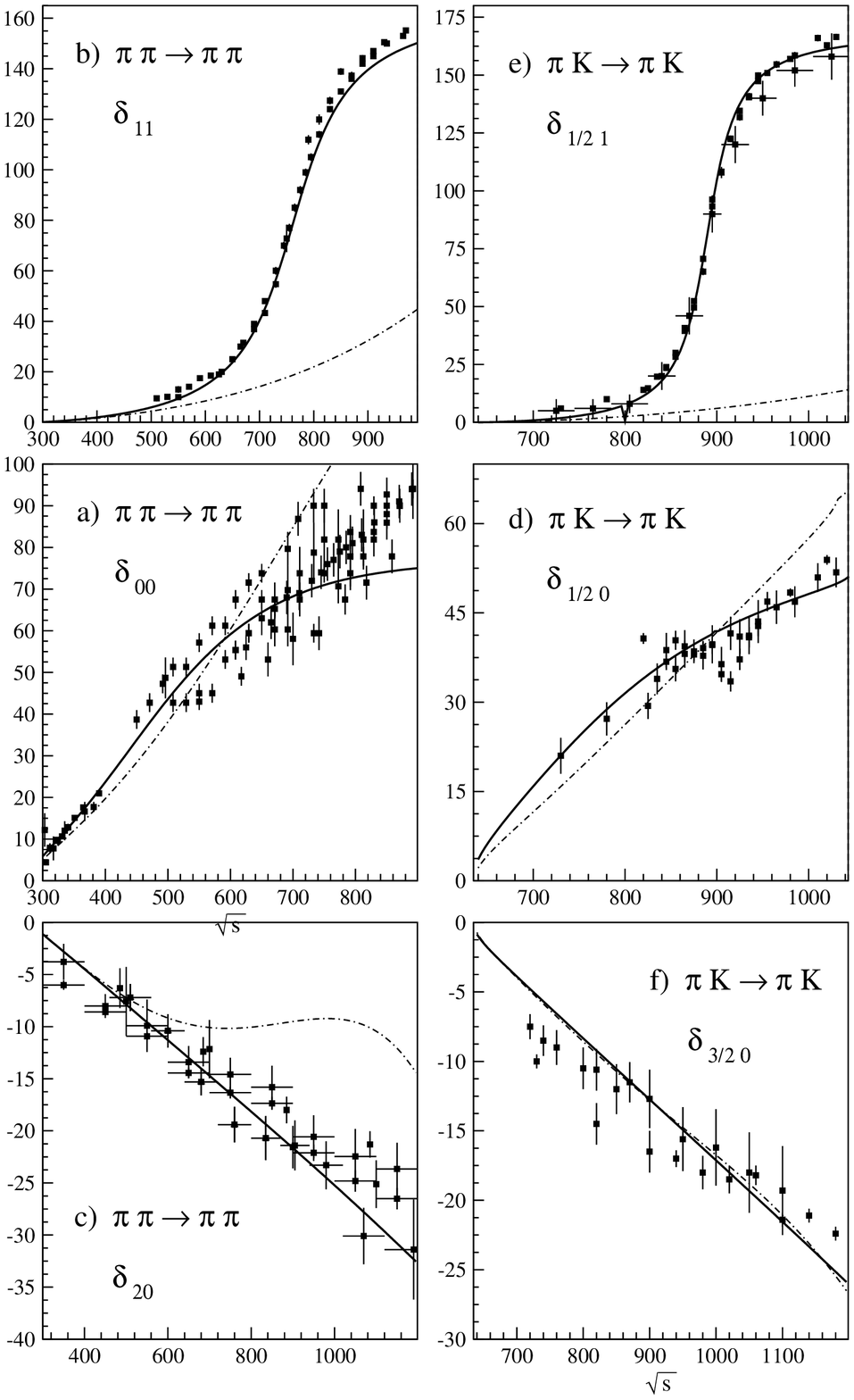}&
  \includegraphics[scale=.48]{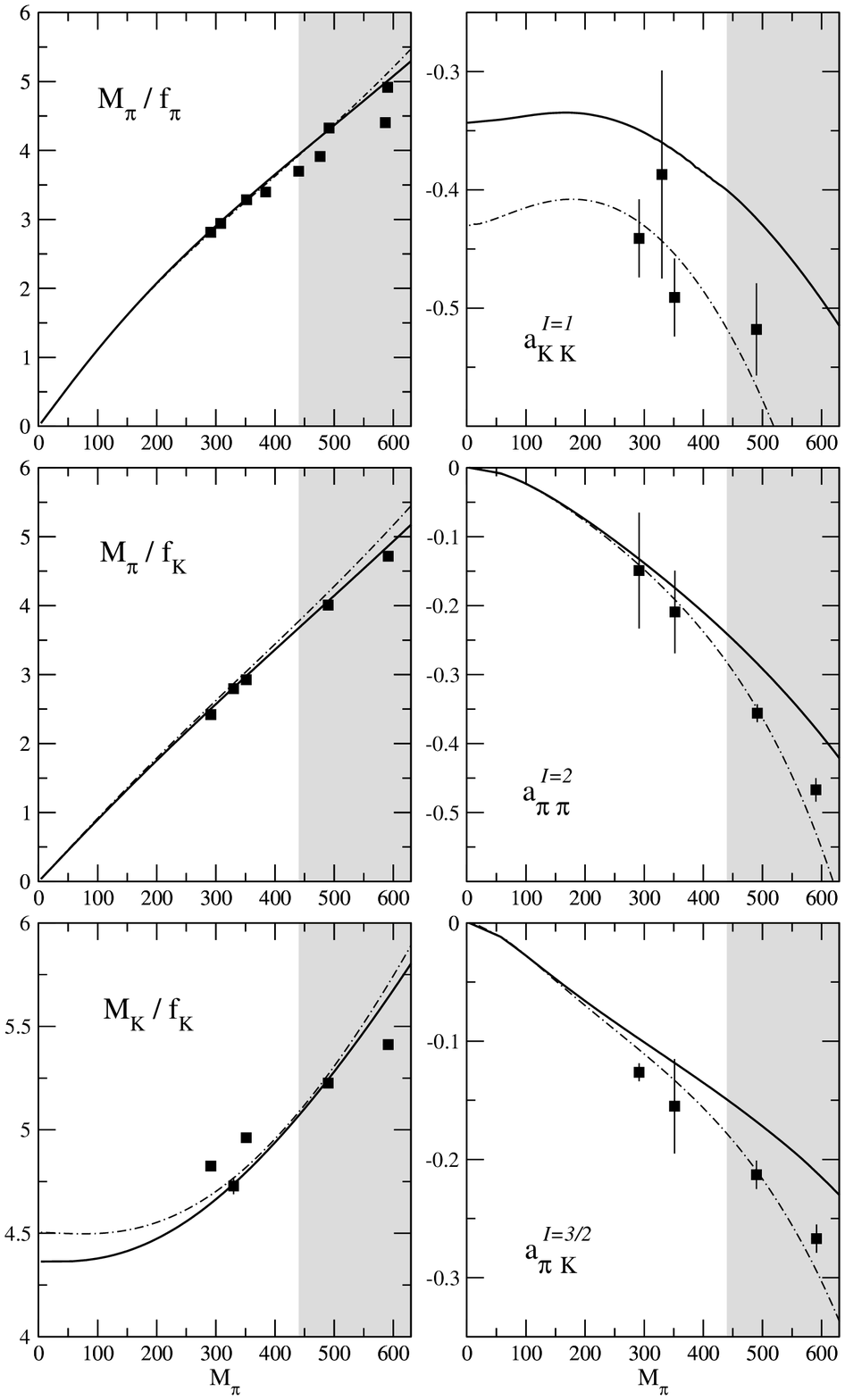}
\end{tabular*}
\vspace*{-.3cm}
\caption{
  Updated IAM fit \cite{Nebreda:2010wv} (continuous line). We also show
non-unitarized ChPT results with the LECs from the $K_{l4}$
two-loop analysis \cite{Amoros:2001cp}(dot-dashed line).  
{\bf Left}: IAM versus data on $\pi\pi$ and $\pi K$
scattering. {\bf Right}: fit results 
compared to lattice calculations  \cite{lattice} on ratios of meson masses and decay constants 
and some scattering lengths. We fit up to $m_\pi=440$ MeV, but even beyond
(grey areas)
lattice results are not described badly. Experimental
references are detailed in \citen{GomezNicola:2001as}.
}
  \label{fig:fits}
\end{figure}

In Fig.\ref{fig:fits} we present some results \cite{Nebreda:2010wv} of an updated fit of the IAM
$\pi\pi$ and $\pi K$ scattering amplitudes to data, simultaneously fitting
the available lattice results on ratios of meson masses and decay constants 
and some scattering lengths.
It is important to remark that the resulting 
LECs are in fairly good agreement
with standard determinations: no fine tuning is required. The $f_0(600)$,
 $\rho(770)$, $\kappa(800)$ and $K^*(892)$ are \emph{not introduced by hand} but 
\emph{generated} as poles in the second 
Riemann sheet of their corresponding partial waves. 
The fact that we do not need 
to model the integrands and 
 {\it the only input parameters are those of ChPT} 
is relevant since we then know
 how to relate our amplitudes to QCD parameters like $N_c$ or $m_q$.



\subsection{Other unitarization techniques within the coupled channel formalism}

Naively one can arrive at \eqref{IAM} in a matrix form, ensuring
coupled channel unitarity, just by expanding
the real part of the inverse $T$ matrix. 
Unfortunately, {\it there is still no dispersive derivation}
including a left cut {\it for the coupled channel case}.
Being much more complicated,  different approximations to $\rea T^ {-1}$
have been used:

$\bullet$ The fully renormalized one-loop ChPT calculation of $\rea T^ {-1}$ 
provides the correct ChPT expansion,
with left cuts approximated to $O(p^4)$ \cite{Guerrero:1998ei,GomezNicola:2001as}. 
Indeed, using
LECs consistent with standard ChPT determinations,
one can describe \cite{GomezNicola:2001as}
below 1.2 GeV all two-body scattering
channels made of pions, kaons or etas. Simultaneously,
this approach \cite{GomezNicola:2001as} generates poles associated to the $\rho(770)$
and $K^*(892)$ vector mesons,
together with the  $f_0(980)$, $a_0(980)$, $f_0(600)$ 
and $\kappa$ (or $K_0(800)$) scalar resonances.

$\bullet$ Originally \cite{Oller:1997ng}, the coupled channel
IAM  was used neglecting crossed loops and tadpoles.
This is considerably 
simpler, and despite the left cut is absent, since its numerical influence
is relatively small, meson-meson data
are described with reasonable LECs 
while generating all poles enumerated above. 
Note that this approximation keeps the s-channel loops
but also the tree level up to $O(p^4)$, which
encodes the effect of  heavier resonances, like the $\rho$. Thus,
contrary to some common belief, the IAM incorporates
the low energy effects of t-channel $\rho$ exchange.

$\bullet$ Finally, if only scalar
 meson-meson scattering is of interest, it is possible to 
use just one cutoff (or another regulator)
that numerically mimics 
the combination of LECs appearing in 
scalar channels. This "`chiral unitary approach"
is very popular, 
even beyond the meson-meson framework,
due to its great simplicity but remarkable success
\cite{Oller:1997ti} and also for its straightforward relation to the Bethe-Salpeter formalism \cite{Nieves:1998hp} that provides  physical insight on unitarization.
With this method it was shown \cite{Oller:2003vf} that, assuming no 
$m_q$ dependence of the cutoff, all light scalar resonances degenerate into an octet and a singlet in the SU(3) limit. Axial-vector mesons have also been generated by using a chiral Lagrangian for the pseudoscalar-vector interaction \cite{Lutz:2003fm}.
\vspace*{-.2cm}

\section{The nature of resonances from their leading $1/N_c$ behavior}
\vspace*{-.2cm}

The QCD $1/N_c$ expansion \cite{'tHooft:1973jz}, valid
 in the whole energy region, 
provides a rigorous definition of $\bar qq$ bound states: their 
masses and widths behave as $O(1)$ and $O(1/N_c)$, respectively.
The QCD leading $1/N_c$ behavior of $f_\pi$
 and the LECs is well known, and ChPT amplitudes 
have no cutoffs or subtraction constants
where spurious $N_c$ dependences could hide.
Hence, by scaling with $N_c$ the ChPT parameters in the IAM, 
the $N_c$ dependence of the mass and width of the resonances has been determined
to one and two loops
 \cite{Pelaez:2003dy,Pelaez:2006nj}.
These are defined from the pole position as $\sqrt{s_{pole}}=M-i\Gamma$.
However, {\it a priori}, one should be careful {\it not to take $N_c$ too large, because the $N_c\to\infty$ limit is a weakly interacting limit}. As shown above, the IAM relies
on the fact that the exact elastic $RC$ contribution dominates the
dispersion relation.
Since the IAM describes data and the resonances within, say, 10 to 20\% errors, this means that at $N_c=3$ the other contributions are not 
approximated badly.  But meson loops, responsible for the $RC$, scale as $3/N_c$ whereas the inaccuracies due to the approximations scale partly as $O(1)$.
Thus, we can estimate that those 10 to 20\% errors at $N_c=3$ become 100\% errors at, say $N_c\sim30$ or $N_c\sim15$, respectively.
Hence we never show results \cite{Pelaez:2003dy,Pelaez:2006nj} beyond $N_c=30$.
Even beyond $N_c\sim15$ they should be interpreted with care.

\begin{figure}[t]
\begin{tabular*}{0.5\textwidth}{@{\extracolsep{-0.cm}}ccc}
      \includegraphics[scale=.46]{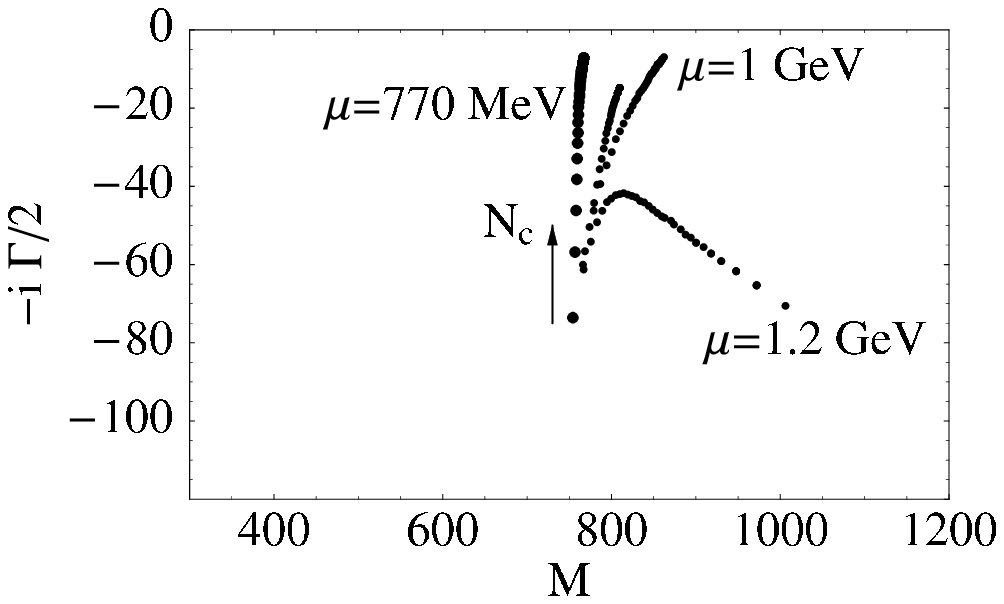}&
      \includegraphics[scale=.43]{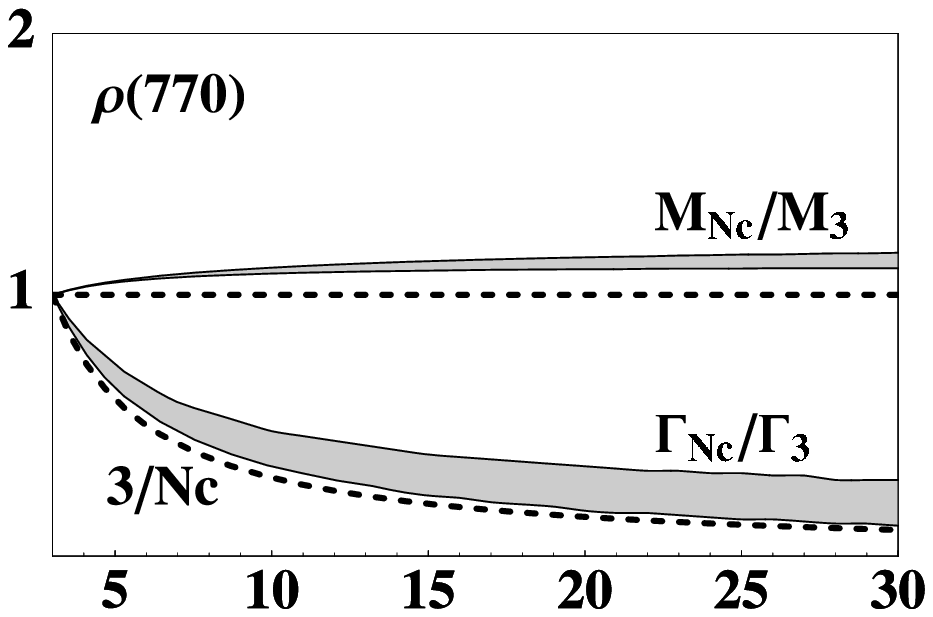}&
      \includegraphics[scale=.43]{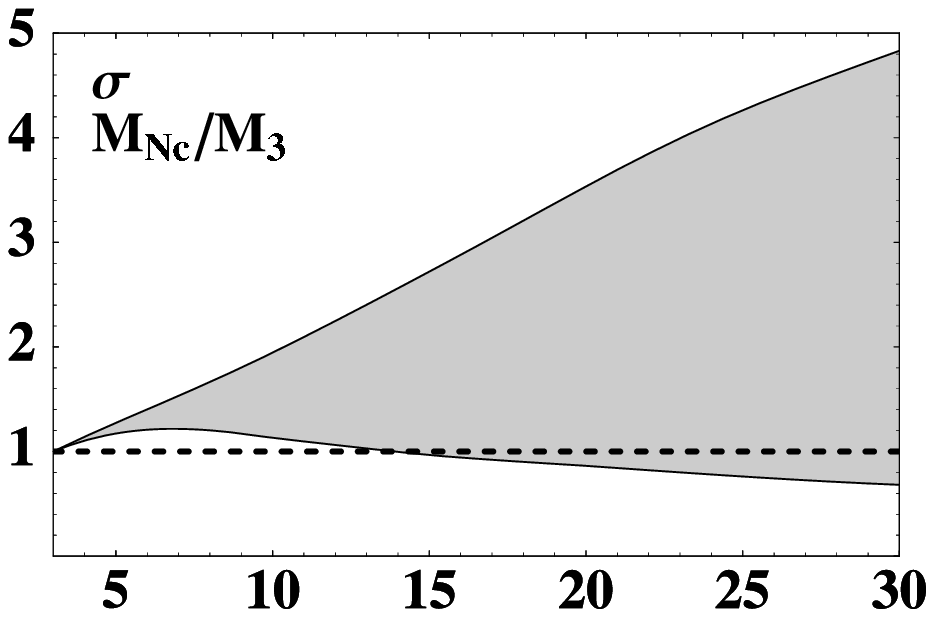}\\
      \includegraphics[scale=.46]{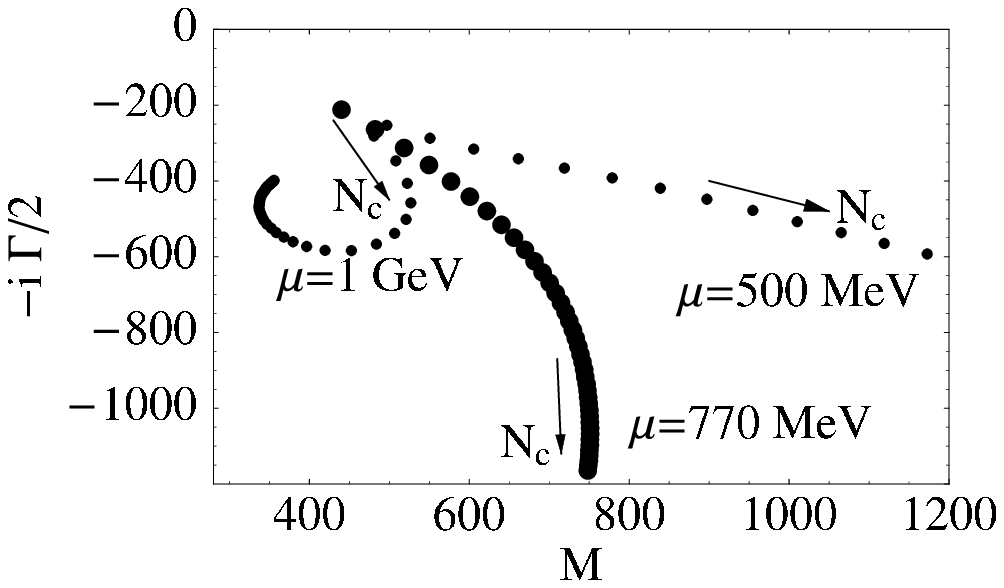}&
      \includegraphics[scale=.43]{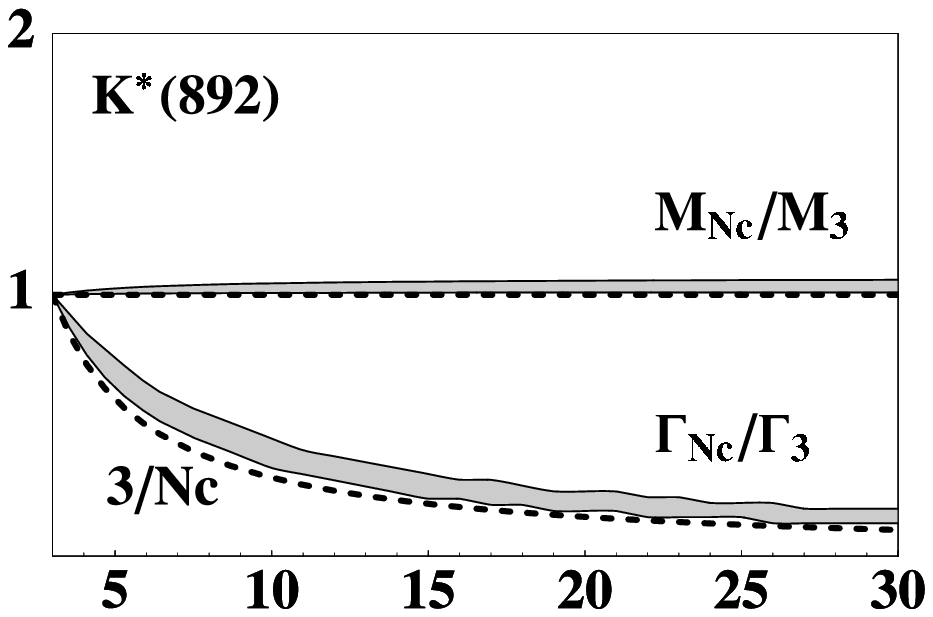}&
      \includegraphics[scale=.43]{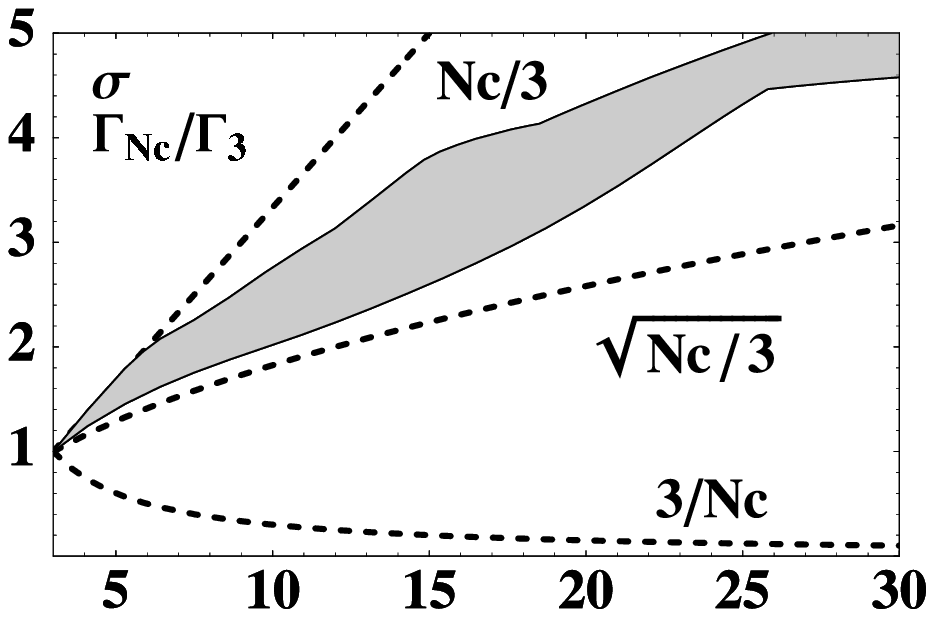}
\end{tabular*}
  \caption{{\bf Left:}  $\rho$ (top) and $\sigma$ (bottom) pole
  trajectories for different values of $\mu$, note that for
  $\mu=1.2$ GeV the $\rho$ pole goes away the real axis. 
 {\bf Center:} $N_c$ behavior of the $\rho$ (top) and $K^*$ (bottom)
  mass and width. {\bf Right:} $N_c$ behavior of the $\sigma$ mass and width. }
  \label{ncSU3}
\end{figure}

Thus, Fig.\ref{ncSU3} shows the
behavior of the $\rho$, $K^*$ and $\sigma$ masses and widths
found in  \cite{Pelaez:2003dy}. The $\rho$ and
$K^*$ neatly follow the expected behavior for 
a $\bar qq$ state: $M\sim 1$, $\Gamma\sim 1/N_c$.
The bands cover the uncertainty $\mu\sim 0.5-1$ GeV
where to apply the $1/Nc$ scaling.
Note also that {\it outside this $\mu$ range}
the $\rho$ meson starts deviating from 
a $\bar qq$ behavior. Something similar occurs to the $K^*(892)$.
Hence, we cannot
apply the $N_c$ scaling at an arbitrary $\mu$ value,
if the well established $\rho$ and $K^*$ $\bar qq$ nature is to be reproduced.



In contrast, the $\sigma$ shows a 
different behavior from that of a pure $\bar qq$:
\emph{near $N_c$=3} both its mass and width
grow with $N_c$, i.e. its pole moves 
away from the real axis. 
Of course,  far from $N_c=3$, and
for some choices of LECs and $\mu$,
the $\sigma$ pole might turn back to the real axis \cite{Pelaez:2006nj,Nieves:2009ez,Sun:2004de}, 
as seen in Fig.2 (top-right). 
But, as commented above, the IAM is less reliable for large $N_c$,
and at most this behavior only suggests that
there {\it might be} a subdominant $\bar q q$ component \cite{Pelaez:2006nj}.
In addition, we have to ensure that the LECs fit data
and reproduce the vector $\bar qq$ behavior.

Since loops are important in determining the scalar pole position,
but are $1/N_c$ suppressed compared to tree level terms with LECs,
we checked the  $O(p^4)$ results
with an $O(p^6)$ IAM calculation in $SU(2)$\cite{Pelaez:2006nj}.
We defined a $\chi^2$-like function to measure how close 
 a resonance is from a $\bar qq$ $N_c$  behavior.
First, we used it at $O(p^4)$ to show
 that it is not possible for the $\sigma$ to behave predominantly as a $\bar q q$
while describing simultaneously the data
and the $\rho$ $\bar qq$ behavior, thus
{\it confirming the robustness of the conclusions for $N_c$ close to 3}. 
Next, we obtained a $O(p^6)$ data fit where
the $\rho$ $\bar qq$ behavior was imposed (see Fig.3, left and center). Note that
both $M_\sigma$ and $\Gamma_\sigma$ grow with $N_c$ near $N_c=3$,
confirming the $O(p^4)$ result of a non $\bar qq$ dominant component.
However, for $N_c$ between 8 and 15, where we still
trust the IAM, $M_\sigma$
becomes constant and $\Gamma_\sigma$ starts decreasing. 
This may hint to a \emph{subdominant $\bar qq$ component},
arising as loops become suppressed when $N_c$ grows.
Finally, by forcing the $\sigma$ to behave 
as a $\bar qq$, 
we found that in the best case (Fig.3, right) this subdominant $\bar qq$
component could become dominant around $N_c>6-8$, at best, but
always with an $N_c\to\infty$ mass  above $\sim 1$ GeV instead of its physical $\sim 450$ MeV value.

\begin{figure}[t]
\hspace*{1cm}
  \hbox{
    \includegraphics[angle=-90,scale=.34]{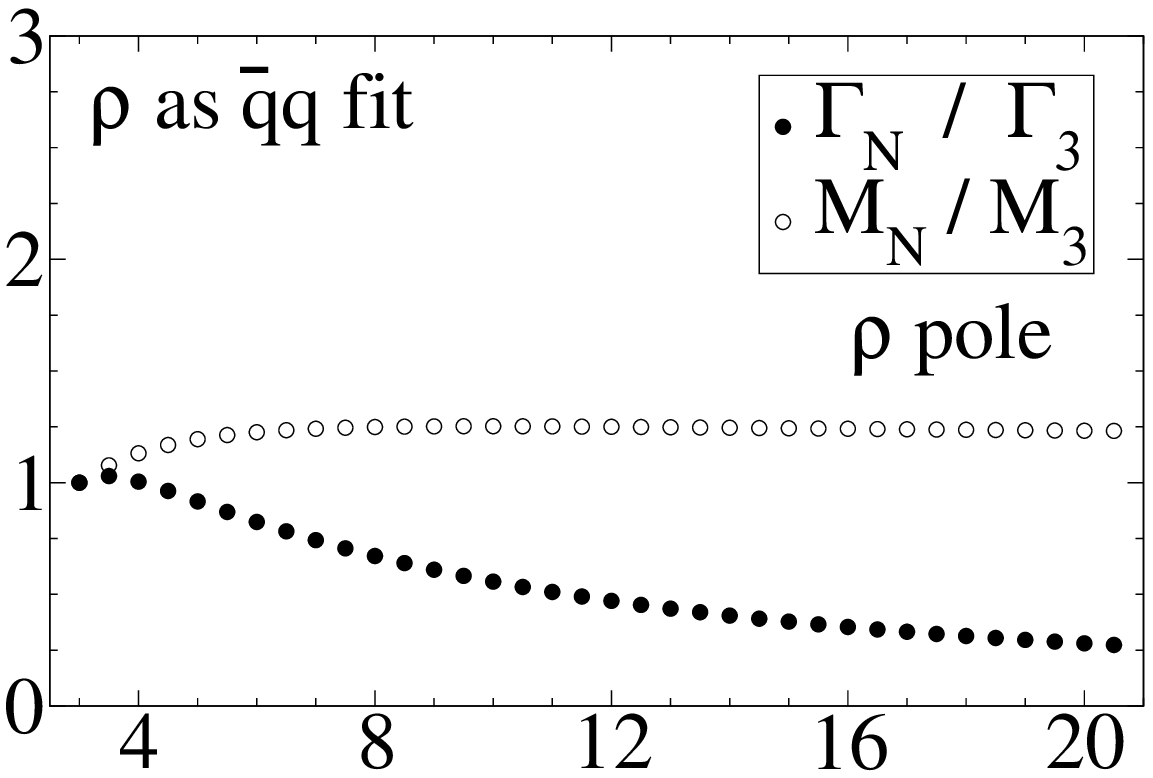}
    \includegraphics[angle=-90,scale=.34]{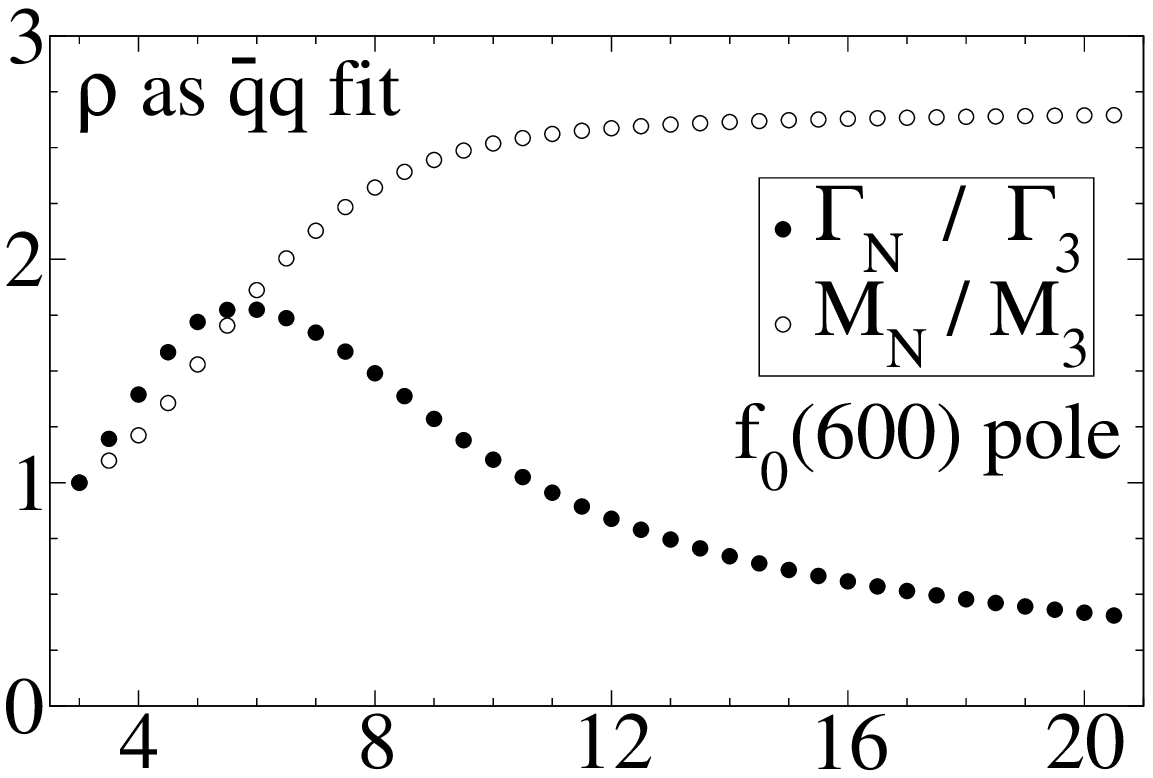}
    \includegraphics[angle=-90,scale=.34]{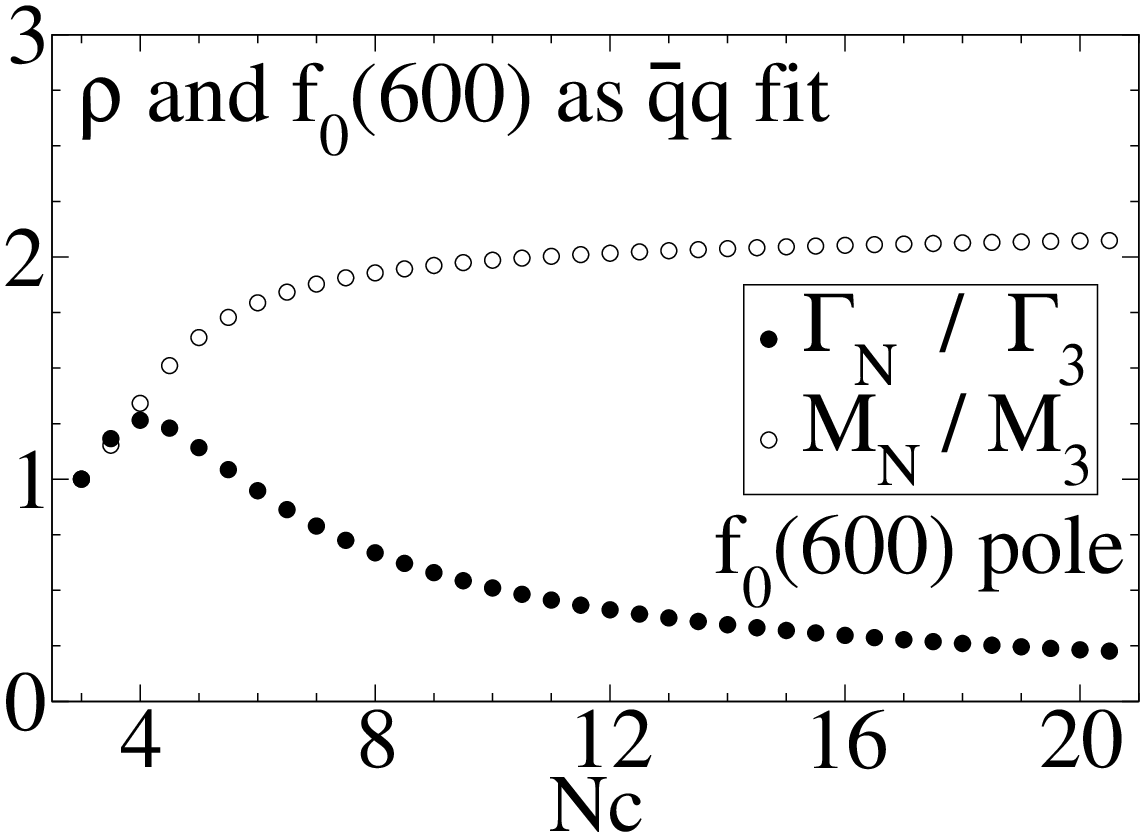}
  }
  \caption{{\bf Left and center}: $N_c$ behavior of the $\rho$
    and $\sigma$ pole at $O(p^6)$ with the ``$\rho$ as 
    $\bar qq$ fit''.
    {\bf Right:} Sigma behavior with $N_c$ at $O(p^6)$ with
    the ``$\rho$ and $\sigma$ as $\bar qq$ fit''.}
  \label{2loops}
\end{figure}

Let us emphasize again \cite{Pelaez:2005fd}
what can and {\it what cannot} be concluded from our results
and clarify some frequent questions and doubts:
\newcounter{input}

$\bullet$ Most likely, scalars are a mixture of different
    states, but 
{\it the \underline{dominant component}  of the $\sigma$ and $\kappa$ {\it in meson-meson scattering} does not behave as a $\bar{q}q$}. If the $\bar{q}q$ was {\it dominant}, they would behave
    as the $\rho$ or the $K^*$ in Fig.2.  {\it However, a smaller fraction of $\bar{q}q$ cannot be
      excluded} and is somewhat favored in our $O(p^6)$ analysis \cite{Pelaez:2006nj}.\\      
$\bullet$ {\it Two meson and some tetraquark states \cite{Jaffe} have a consistent
      ``qualitative'' behavior}, i.e., both disappear in
    the  meson-meson scattering continuum as $N_c$
    increases. Our results are not able yet
     to establish the nature of that dominant component.
To do so other tools\cite{Weinberg,baru} might be necessary.
The most we can state is that the behavior of
      two-meson states or some tetraquarks might be
qualitatively consistent.

The $N_c\rightarrow\infty$
limit has been studied in \citen{Sun:2004de,Nieves:2009ez}. Apart from its
mathematical interest, it could have some physical
relevance if the data and the large $N_c$ uncertainty on the
choice of scale were more accurate. Nevertheless:\\
$\bullet$ {\it A priori the IAM is not reliable in the  $N_c\rightarrow\infty$ limit}, since that is a weakly interacting theory, where 
exact unitarity
becomes less relevant in confront of other approximations made in the IAM derivation. It has been shown \cite{Nieves:2009ez} that it might work
well in that limit in the vector channel of QCD but not in the scalar channel.

$\bullet$ Another reason to keep $N_c$ not too
far from 3 is that we have not included the $\eta'(980)$,
whose mass is related to the $U_A(1)$ anomaly and scales as 
$\sqrt{3/N_c}$.
Nevertheless, if in our calculations we keep $N_c<30$, its mass
would be $>310\,$MeV and thus pions are still the only relevant degrees
of freedom in the $\sigma$ region.

$\bullet$ {\it Contrary to the leading $1/N_c$ behavior 
\underline{in the vicinity of  $N_c=3$},
the $N_c\rightarrow\infty$ 
limit does not give information on the ``dominant component''
of light scalars.} The reason was 
commented above: in contrast to $\bar{q}q$
states, that become bound, 
 two-meson and some tetraquark
states dissolve in the 
continuum as $N_c\rightarrow\infty$. 
Thus, even if we started with an infinitesimal $\bar{q}q$ component
in a resonance, for a sufficiently large $N_c$ 
it may become dominant, and beyond 
that $N_c$ the associated pole
would behave as a $\bar{q}q$ state.
Also, since the mixings of 
different components could change
with $N_c$, a too large $N_c$ could alter significantly
the original mixings.

Actually, this is what happens for the one-loop IAM $\sigma$ resonance 
for $N_c\to\infty$, but 
it does {\it not} necessarily mean that 
the ``correct interpretation [...] is that
the $\sigma$ pole is a conventional 
$\bar{q}q$ meson environed by heavy pion clouds'' \cite{Sun:2004de}.
That the $\sigma$ is not conventional is simply seen by comparing
it with the ``conventional'' $\rho$ and $K^*$ in Fig. 2.
A large two-meson component is consistent, but so is a tetraquark. Actually,
the $N_c\rightarrow\infty$ of the one-loop unitarized ChPT 
pole in the scalar channel
 limit is not unique \cite{Sun:2004de,Nieves:2009ez} 
given the uncertainty
in the chiral parameters. Moreover, despite the  one-loop IAM could make
sense in the $N_c\rightarrow\infty$ limit for the vector channel\cite{Nieves:2009ez},
in the scalar channel 
it can lead to phenomenological inconsistencies 
\cite{Nieves:2009ez} for some LECs, since poles can even move to negative 
 squared mass values (weird), 
to infinity or to a  positive mass square. 
Hence, robust 
conclusions on the dominant light scalar component
can be obtained not too far
 from real life, say $N_c<15$ or 30,  for a $\mu$ choice between
roughly $0.5$ and 1 GeV, that simultaneously ensures the $\bar qq$ dependence for the $\rho$ and $K^*$ mesons.  Note, however, that under these same 
conditions the two-loop IAM still finds, not only 
a dominant non-$\bar qq$ component, 
but also a hint of a $\bar qq$ subdominant component\cite{Pelaez:2006nj}, which
 is not conventional in the sense that it appears 
at a much higher mass than the physical $\sigma$. This subdominant component at that higher mass seems
to be needed to ensure fulfillment of local duality\cite{deElvira:2010an} for $Nc>3$.
This may support the existence of a second scalar octet, a $\bar qq$ now,
  above 1 GeV \cite{VanBeveren:1986ea}.  

Finally, using not the IAM, but the chiral unitary approach with a natural 
range for the cutoff $N_c$ dependence, it has also been suggested \cite{Geng:2008ag} that a 
large, in some cases dominant, non $\bar qq$ behavior could exist in axial vector mesons.
\vspace*{-.2cm}

\section{Quark mass dependence of resonances}
\vspace*{-.2cm}

ChPT provides a rigorous expansion of meson masses in terms of $m_q$
(at leading order $M_{meson}^2\sim m_q$). Thus, 
by changing the meson masses in the amplitudes, we see how the poles
generated with the IAM depend on $m_q$.
We report here the SU(2) analysis\cite{Hanhart:2008mx} of $\rho$ and $\sigma$ as well
the SU(3) analysis\cite{Nebreda:2010wv} of non-strange, $\rho$ and $\sigma$, 
and strange, $\kappa(800)$ and $K^*(892)$, resonances.

The values of $m_\pi$ considered should fall within the ChPT
range of applicability and allow for some elastic $\pi\pi$ and $\pi K$
regime below $K\bar K$ or $K\eta$ thresholds, respectively. Both criteria are
satisfied if $m_\pi\leq 440$ MeV, since $SU(3)$ ChPT
still works with such kaon masses, and because for
$m_\pi\simeq 440$ MeV, the kaon mass becomes $\simeq 600$ MeV. 
Of course, we expect higher order
corrections, which are not considered here, to become more
relevant as $m_\pi$ is increased. Thus, our results become less
reliable as $m_\pi$ increases due to the $O(p^6)$ 
corrections which we have neglected.

\begin{figure}[t]
   \centering
     \hbox{
       \includegraphics[scale=0.25,angle=0]{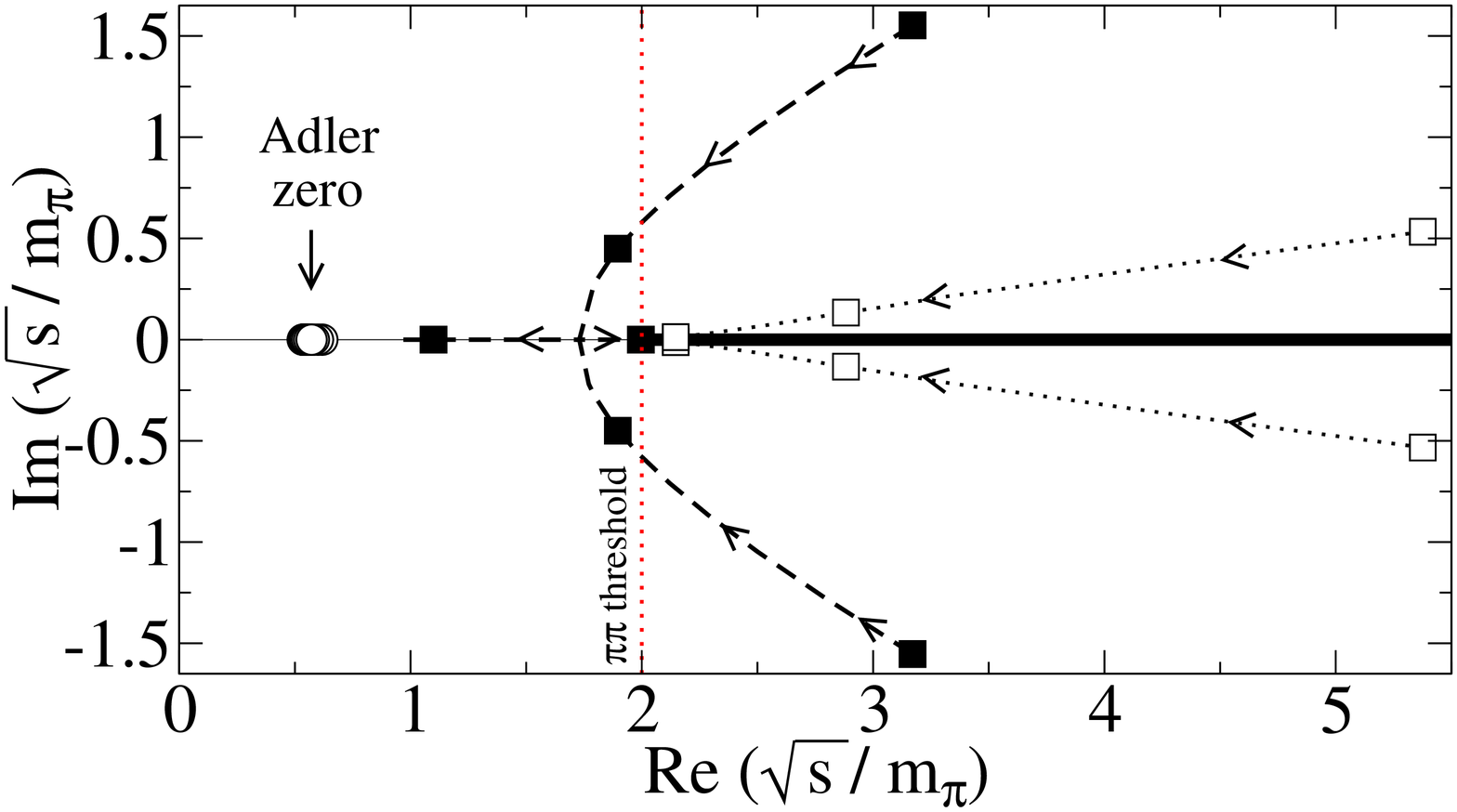}
       \includegraphics[scale=0.57,angle=0]{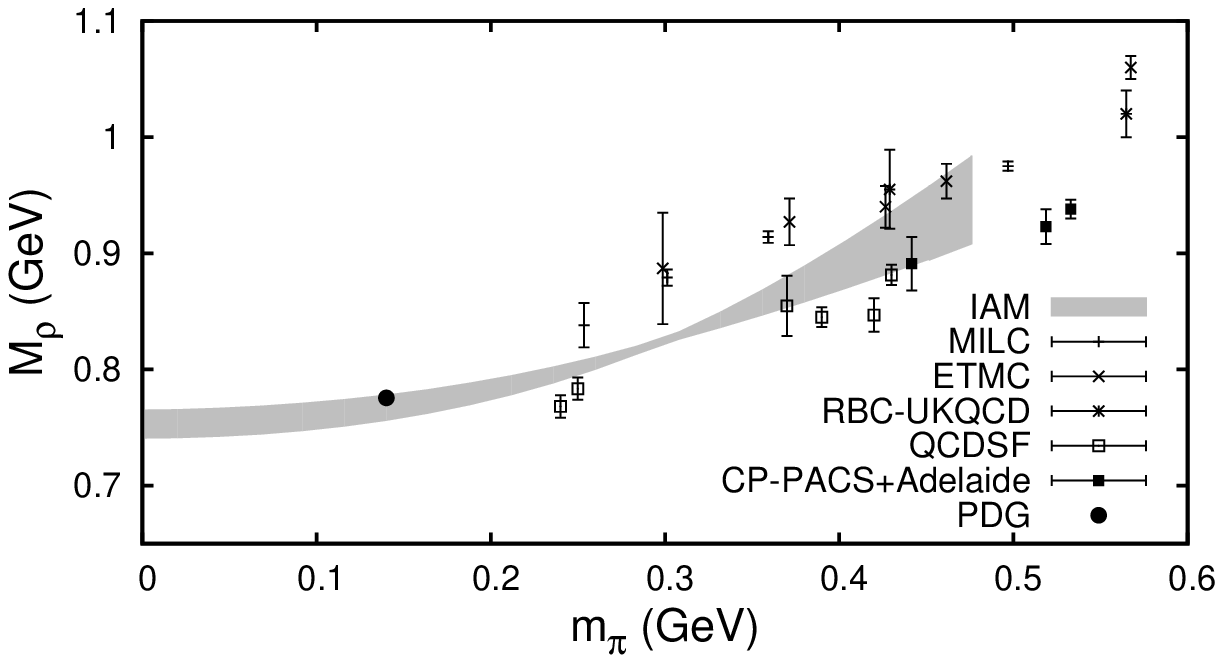}
     }
     \caption{ {\bf Left:} Movement of the $\sigma$ (dashed lines) 
       and $\rho$ (dotted
       lines) poles for increasing $m_\pi$ (direction indicated by the
       arrows) on the second sheet.  The filled (open) boxes denote the
       pole positions for the $\sigma$ ($\rho$) at pion masses $m_\pi=1,\
       2,$ and $3 \times m_\pi^{\rm phys}$, respectively. For
       $m_\pi=3m_\pi^{\rm phys}$ three poles accumulate 
       very near the  threshold. All poles are
       always far enough from the Adler zero (circles).
       {\bf Right:} Comparison of our results for the $M_\rho$
       dependence on $m_\pi$ with some recent lattice results\cite{lattice1}. 
       The grey band covers the
       error coming from the LECs uncertainties.}
       \label{poles}
 \end{figure}

Fig. \ref{poles} (left) shows the evolution of the $\sigma$ and $\rho$
pole positions as $m_\pi$ is increased. In order to see the pole
movements relative to the two pion threshold, which is also increasing,
we use units of $m_\pi$, so the threshold is
fixed at $\sqrt{s}=2$. Both poles move closer to threshold and
they approach the real axis. The $\rho$ poles reach the real axis
at the same time that they cross threshold.
One of them jumps into the first sheet and stays below
threshold in the real axis as a bound state, while its conjugate
partner remains on the second sheet practically at the very same
position as that in the first. In contrast, the $\sigma$
poles go below threshold with a finite imaginary part before they
meet in the real axis, still on the second sheet, becoming
virtual states. As $m_\pi$ increases, one pole
moves toward threshold and jumps through the branch point to the
first sheet staying in the real axis below threshold, very
close to it as $m_\pi$ keeps growing. The other $\sigma$ pole moves
down in energies away from threshold and remains 
on the second sheet.
These  very asymmetric poles could
signal a prominent molecular component \cite{Weinberg,baru},
at least for large pion masses.
Similar movements were found within quark models
\cite{vanBeveren:2002gy} and a finite density analysis 
\cite{FernandezFraile:2007fv}.

Fig. \ref{poles} (right) shows our results for the $\rho$ mass
dependence on $m_\pi$ compared with some recent lattice 
results~\cite{lattice1}, and the
PDG value for the $\rho$ mass. 
Now the mass is defined as the point where the phase shift crosses $\pi/2$,
except for those $m_\pi$ values where the $\rho$ becomes a bound
state, where it is defined again from the pole position.
Taking into
account the incompatibilities between
different lattice collaborations, we find a qualitative good
agreement with lattice results. Note also that the $m_\pi$ dependence
in our approach is correct only up to NLO in ChPT, and
we expect higher order corrections to be important for
large pion masses. The $M_\rho$ dependence on $m_\pi$
 agrees also with estimations 
for the two first coefficients of its chiral expansion \cite{bruns}.

\begin{figure}[t]
\begin{tabular}{@{\extracolsep{-0.4cm}}cccc}
  \includegraphics[scale=.28,angle=-90]{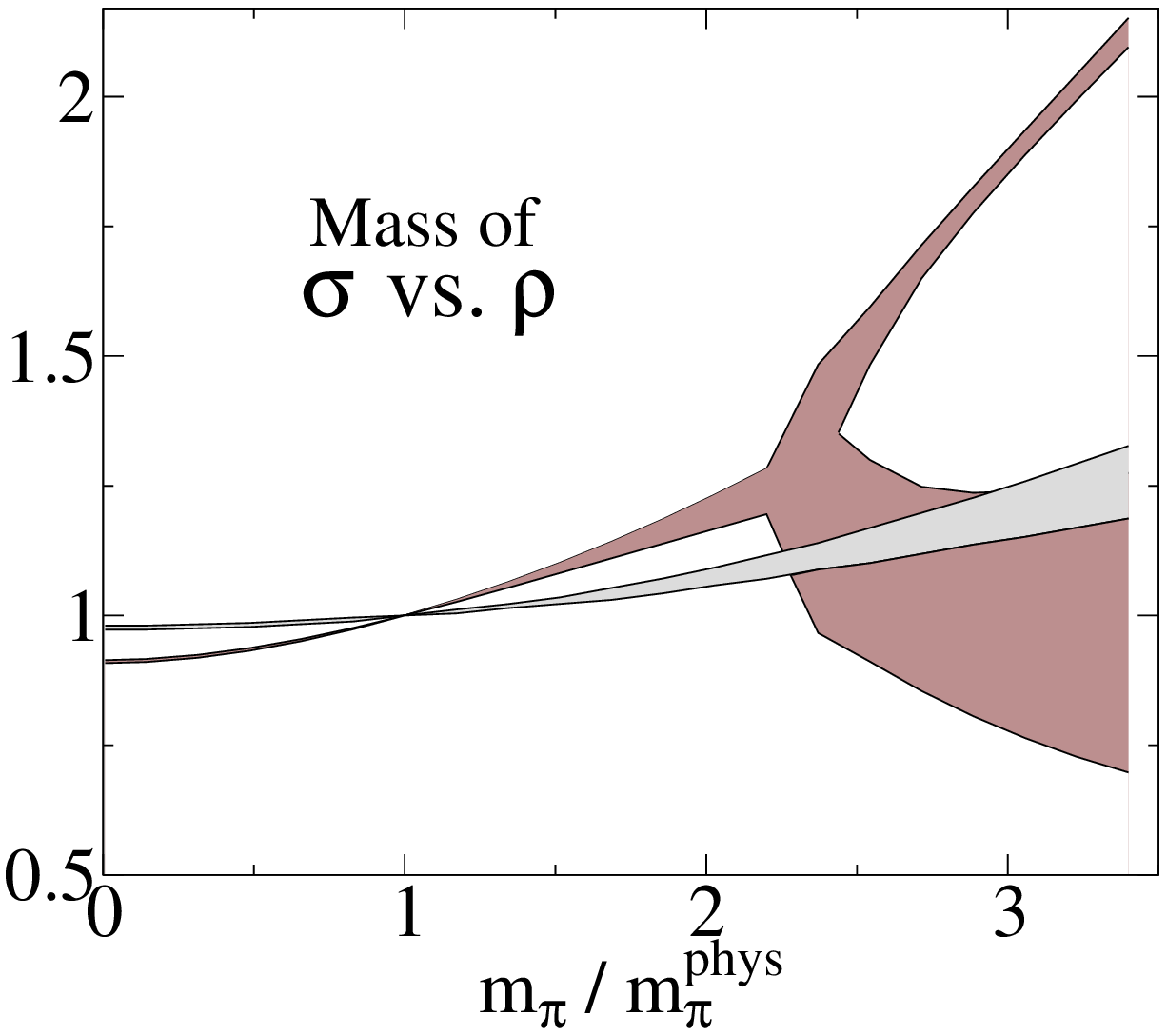}&
  \includegraphics[scale=.28,angle=-90]{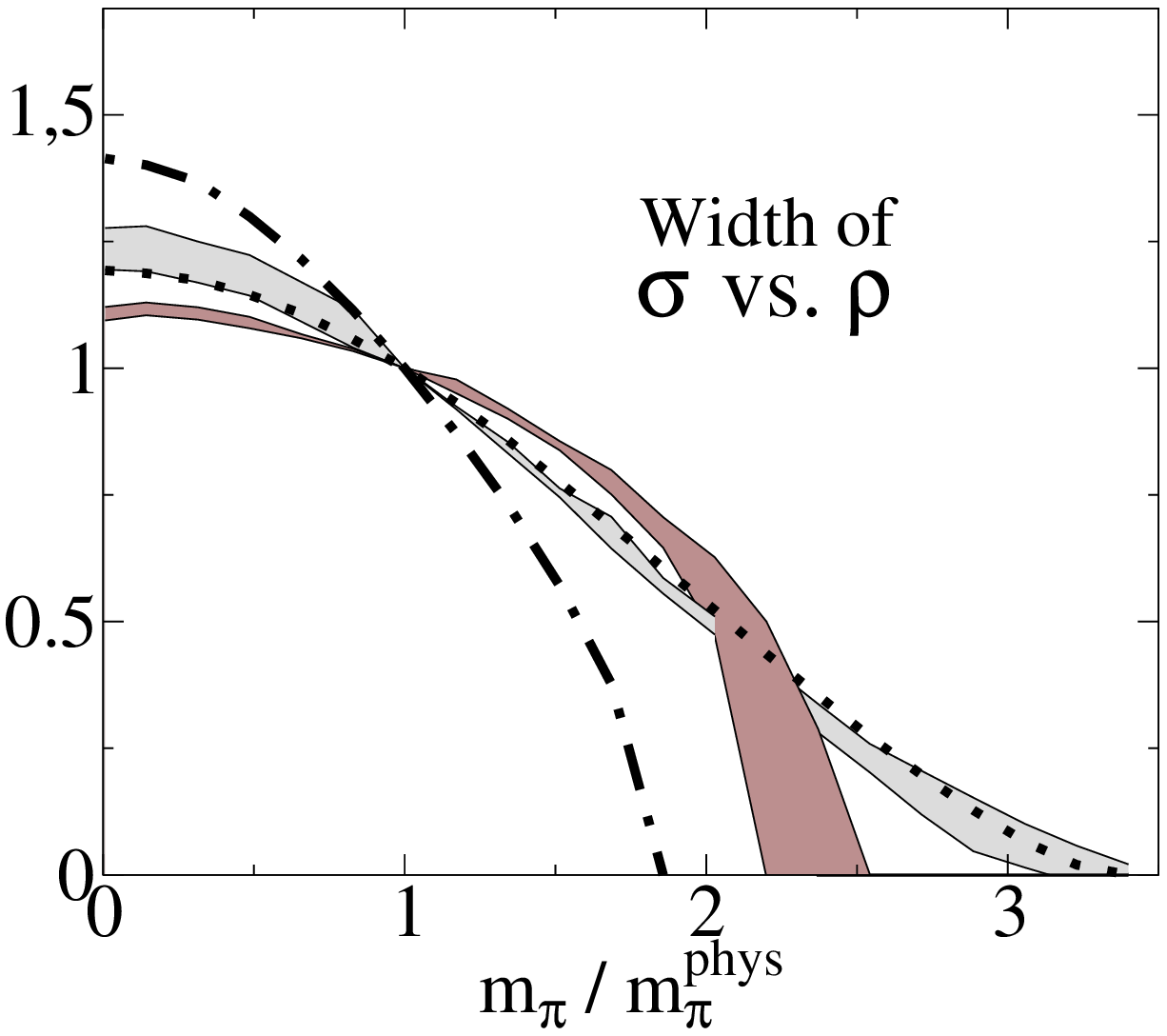}&
  \includegraphics[scale=.28,angle=-90]{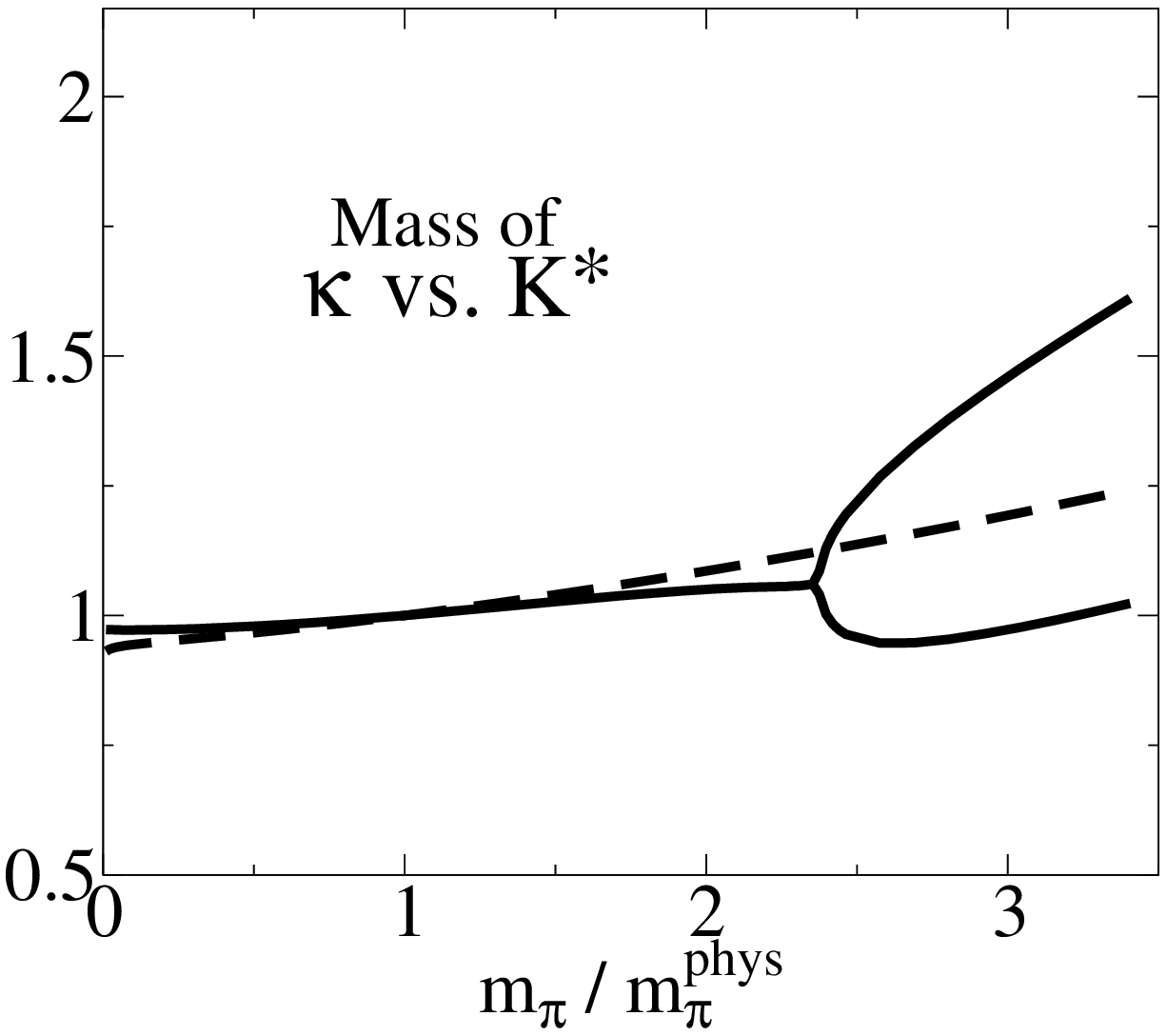}&
  \includegraphics[scale=.28,angle=-90]{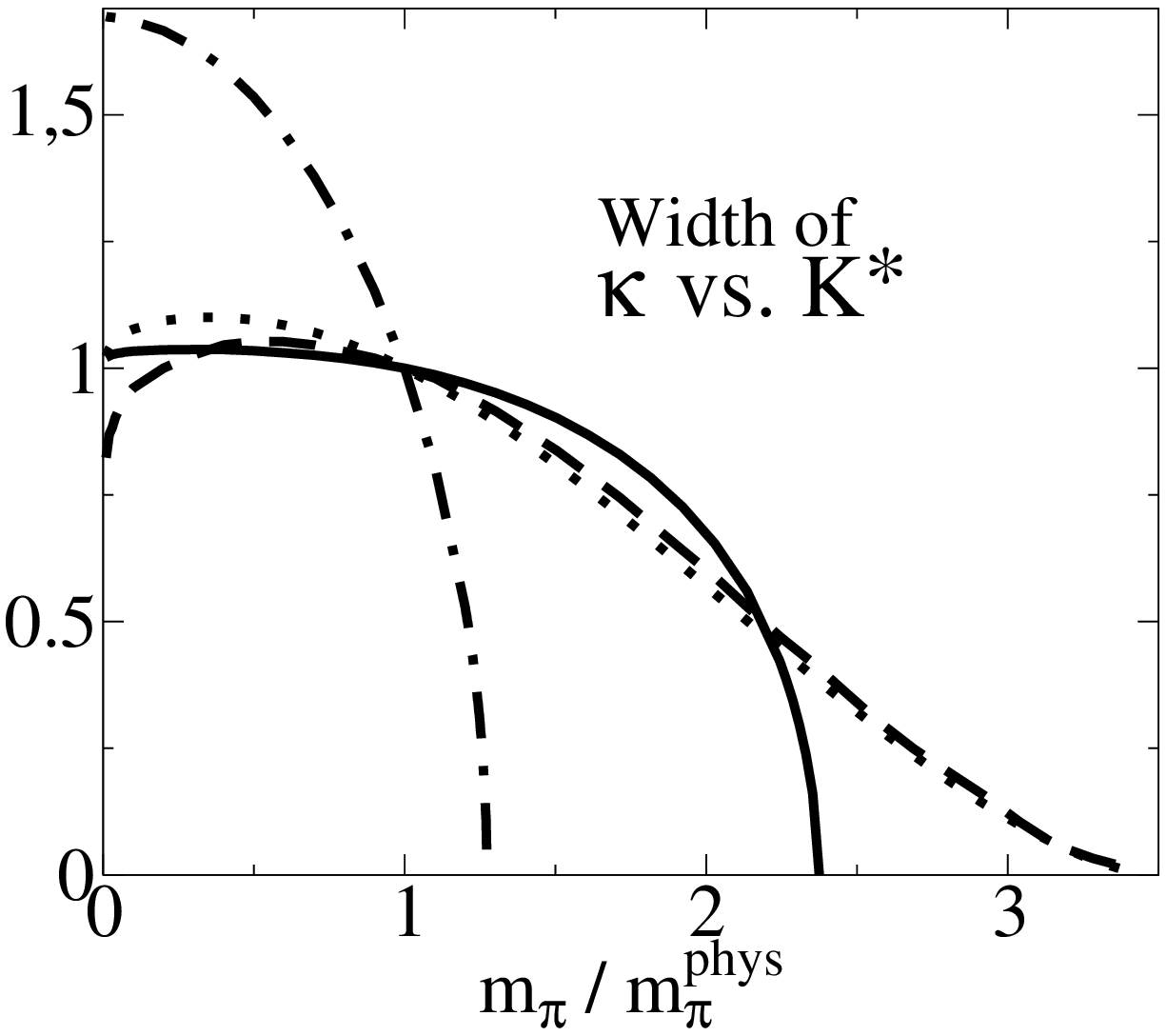}
  \end{tabular}
 \caption{\label{massandwidth} $m_\pi$ dependence of resonance masses
       and widths in units of the physical values. 
       In the two left panels the dark (light) band shows the results for
       the $\sigma$ ($\rho$). The band width reflects
       the uncertainties in the SU(2) LECs.  
Similarly, the two right panels, calculated within SU(3)
\cite{Nebreda:2010wv}, show the behavior for the 
$K^*(892)$ (continuous) and $\kappa(800)$ (dashed).
The (dotted) dot-dashed line shows the $m_\pi$ dependence of the 
corresponding vector (scalar) width from the change of phase space only, assuming
       a constant coupling of the resonance to two mesons.
}
 \end{figure}

In Fig. \ref{massandwidth} (left) we compare the $m_\pi$ dependence
of $M_\rho$ and $M_\sigma$ (defined from the pole position
$\sqrt{s_{pole}}=M-i\Gamma /2$), normalized to their physical values.
The bands cover the LECs uncertainties. Both masses
grow with $m_\pi$, but $M_\sigma$ grows faster than $M_\rho$. 
Below $m_\pi\simeq 2.4 \, m_\pi^{\rm phys} $ we only show one line because the two
conjugate $\sigma$ poles have the same mass. Above $2.4 \, m_\pi^{\rm phys} $, these
two poles lie on the real axis with two different masses. The
heavier pole goes towards threshold and around $m_\pi\simeq 3.3 \, m_\pi^{\rm phys}$ moves into the
first sheet, but that is beyond our applicability limit.

In the next panel of Fig. \ref{massandwidth} we compare the $m_\pi$
dependence of $\Gamma_\rho$ and $\Gamma_\sigma$ normalized to their
physical values: note that both widths become smaller. 
We compare this decrease with the expected phase space reduction 
 as resonances approach the $\pi\pi$ threshold.
We find that $\Gamma_\rho$ follows very well
this expected behavior, which
implies that the $\rho\pi\pi$ coupling is almost $m_\pi$ independent.
In contrast, $\Gamma_\sigma$ deviates from the
phase space reduction expectation. This suggests a strong $m_\pi$
dependence of the $\sigma$ coupling to two pions, necessarily
present for molecular states \cite{baru,mol}.

Finally, in the last two panels of Fig.\ref{massandwidth}
we compare the mass and width dependence on $\hat m$
of the $\kappa(800)$ versus
the $K^*(892)$, keeping $m_s$ fixed \cite{Nebreda:2010wv}. 
Note that the same pattern of the $\sigma-\rho$ system
is repeated. Belonging to the same octet, $K^*(892)$ and $\rho$ behave very similarly, and 
both their widths follow just phase space reduction. 
The  $\sigma$  and $\kappa$ behaviors are 
only qualitatively similar,
the latter being somewhat softer. This might be partly due to a possible significant 
admixture of singlet state in the $\sigma$.
The dependence of these resonances on $m_s$ has been also studied in 
Ref.\citen{Nebreda:2010wv}.
 
\vspace*{-.3cm}

\section{Summary}
\vspace*{-.2cm}

We have reviewed how the Inverse Amplitude Method (IAM)
\cite{GomezNicola:2007qj} is derived from the first principles of
analyticity, unitarity, and Chiral Perturbation Theory (ChPT)  at low energies. 
It is able to generate, as
poles in the amplitudes,
the light resonances appearing in meson-meson elastic scattering, 
 without any a priori assumptions. Up to a given order in ChPT, it
yields the correct dependences on $\hat m, m_s$ and $N_c$. 

The leading $1/N_c$ behavior suggests 
that the dominant component of light scalars
does not behave as a
$\bar{q}q$ state as $N_c$ increases not far from $N_c=3$.
When using the two loop IAM
result in SU(2), below $N_c\sim\,$15 or 30,  
there is a hint of a subdominant $\bar{q}q$ component, but
 arising at roughly twice the mass of the physical $\sigma$.

We have studied the pion (quark) mass dependence of the
$f_0(600), \rho(770), \kappa(800)$ and $K^*(892)$
poles \cite{Hanhart:2008mx,Nebreda:2010wv} and 
how
they become bound states: softly for vectors and 
with a non-analyticity for scalars. 
We found that the
vector-meson-meson coupling constant is almost $m_\pi$ independent and 
a qualitative agreement with some lattice results for the
$\rho$ mass evolution with $m_\pi$.  These results may  be
relevant for studies of the meson spectrum\cite{Prelovsek:2010kg} and form factors\cite{Guo:2008nc} on the lattice. 

\section*{Acknowledgments}
\vspace*{-.2cm}
Work partially supported by Spanish MICINN: FPA2007-29115-E,
FPA2008-00592 and FIS2006-03438,
U.Complutense/ Banco Santander grant PR34/07-15875-BSCH and
UCM-BSCH GR58/08 910309 and the EU-Research Infrastructure
Integrating Activity
``Study of Strongly Interacting Matter''
(HadronPhysics2, Grant n 227431)
under the EU Seventh Framework Programme.


\begin{thebibliography}{99}
  

  \bibitem{chpt1}
    S. Weinberg, Physica {\bf A96} (1979) 327.
    J.~Gasser and H.~Leutwyler,
    Annals Phys.\  {\bf 158} (1984) 142;
Nucl.\ Phys.\ B {\bf 250} (1985) 465.


\bibitem{'tHooft:1973jz}
G.~'t Hooft,
Nucl.\ Phys.\ B {\bf 72} (1974) 461.
E.~Witten,
Ann. Phys.\  {\bf 128} (1980) 363.


\bibitem{gilberto}
 I.~Caprini et al.,
  Phys.\ Rev.\ Lett.\  {\bf 96} (2006) 132001.
  S.~Descotes-Genon and B.~Moussallam,
  Eur.\ Phys.\ J.\  C {\bf 48}, 553 (2006)


\bibitem{Truong:1988zp}
  T.~N.~Truong,
  Phys.\ Rev.\ Lett.\  {\bf 61}, 2526 (1988).
  T.~N.~Truong,
  Phys.\ Rev.\ Lett.\  {\bf 67}, 2260 (1991).
  A.~Dobado, M.~J.~Herrero and T.~N.~Truong,
  Phys.\ Lett.\  B {\bf 235}, 134 (1990).


\bibitem{Dobado:1992ha}
A.~Dobado and J.~R.~Pel\'aez,
Phys.\ Rev.\ D {\bf 47} (1993) 4883;
Phys.\ Rev.\ D {\bf 56} (1997) 3057.




\bibitem{GomezNicola:2007qj}
  A.~Gomez Nicola, J.~R.~Pelaez and G.~Rios,
  Phys.\ Rev.\  D {\bf 77}, 056006 (2008)




 \bibitem{Gasser:1990bv}
  J.~Gasser and U.-G.~Mei\ss ner,
  Nucl.\ Phys.\  B {\bf 357} (1991) 90.


\bibitem{Nebreda:2010wv}
  J.~Nebreda and J.~R.~Pelaez.,
  Phys.\ Rev.\  D {\bf 81}, 054035 (2010).

\bibitem{Amoros:2001cp}
  G.~Amoros, J.~Bijnens and P.~Talavera,
  Nucl.\ Phys.\  B {\bf 602}, 87 (2001)

\bibitem{GomezNicola:2001as}
  A.~Gomez Nicola and J.~R.~Pelaez,
  Phys.\ Rev.\ D {\bf 65}, 054009 (2002).
AIP Conf.\ Proc.\  {\bf 660} (2003) 102.
[hep-ph/0301049].
  J.~R.~Pelaez,
  Mod.\ Phys.\ Lett.\ A {\bf 19}, 2879 (2004)



\bibitem{lattice}
  S.~R.~Beane  {\it et al.} [NPLQCD Collaboration],
  Phys.\ Rev.\  D {\bf 77}, 094507 (2008) and
  Phys.\ Rev.\  D {\bf 77}, 014505 (2008);
  S.~R.~Beane {\it et al.}
  Phys.\ Rev.\  D {\bf 74}, 114503 (2006);
  Ph.~Boucaud {\it et al.}  [ETM collaboration],
  Comput.\ Phys.\ Commun.\  {\bf 179}, 695 (2008).
  



\bibitem{Guerrero:1998ei}
  F.~Guerrero and J.~A.~Oller,
  Nucl.\ Phys.\ B {\bf 537}, 459 (1999)
  [Erratum-ibid.\ B {\bf 602}, 641 (2001)]


\bibitem{Oller:1997ng}
J.~A.~Oller, E.~Oset and J.~R.~Pelaez,
Phys.\ Rev.\ Lett.\  {\bf 80} (1998) 3452;
Phys.\ Rev.\ D {\bf 59} (1999) 074001

\bibitem{Oller:1997ti}
  J.~A.~Oller and E.~Oset,
  Nucl.\ Phys.\ A {\bf 620}, 438 (1997)
  [Erratum-ibid.\ A {\bf 652}, 407 (1999)]
 and Phys.\ Rev.\ D {\bf 62} (2000) 114017.


\bibitem{Nieves:1998hp}
  J.~Nieves and E.~Ruiz Arriola,
  Phys.\ Lett.\  B {\bf 455}, 30 (1999)

\bibitem{Oller:2003vf}
  J.~A.~Oller,
  Nucl.\ Phys.\  A {\bf 727}, 353 (2003)


\bibitem{Lutz:2003fm}
M.~F.~M.~Lutz and E.~E.~Kolomeitsev,
Nucl.\ Phys.\  A {\bf 730}, 392 (2004).
 L.~Roca, E.~Oset and J.~Singh,
 Phys.\ Rev.\  D {\bf 72}, 014002 (2005).
 L.~S.~Geng, E.~Oset, L.~Roca and J.~A.~Oller,
 Phys.\ Rev.\  D {\bf 75}, 014017 (2007).



\bibitem{Pelaez:2003dy}
  J.~R.~Pelaez,
  Phys.\ Rev.\ Lett.\  {\bf 92}, 102001 (2004). 


\bibitem{Pelaez:2006nj}
  J.~R.~Pelaez and G.~Rios,
  Phys.\ Rev.\ Lett.\  {\bf 97}, 242002 (2006)
  
\bibitem{Nieves:2009ez}
  J.~Nieves and E.~R.~Arriola,
  Phys.\ Rev.\  D {\bf 80}, 045023 (2009)

\bibitem{Sun:2004de}
  Z.~X.~Sun, {\it et al.}
  [arXiv:hep-ph/0411375] and
  Z.~X.~Sun, {\it et al.}
  [arXiv:hep-ph/0503195].


\bibitem{Pelaez:2005fd}
  J.~R.~Pelaez,
  arXiv:hep-ph/0509284. Proceedings of the 11th International Conference on Elastic and Diffractive Scattering,  Blois, France, 15-20 May 2005.
  J.~R.~Pelaez and G.~Rios,
  arXiv:0905.4689 [hep-ph].
Proceedings of Excited QCD, Zakopane, Poland, 8-14 Feb 2009. 



\bibitem{Jaffe} R. L. Jaffe, Proc. of the Intl. Symposium
on Lepton and Photon Interactions at High Energies. Physikalisches Institut, Univ. of Bonn (1981). ISBN: 3-9800625-0-3.  

\bibitem{deElvira:2010an}
  J.~R.~de Elvira, J.~R.~Pelaez, M.~R.~Pennington and D.~J.~Wilson,
  arXiv:1001.2746 [hep-ph].


\bibitem{VanBeveren:1986ea}
  E.~Van Beveren, {\it et al.} 
  Z.\ Phys.\ C {\bf 30}, 615 (1986)
and 
hep-ph/0606022.
E.~van Beveren and G.~Rupp,
Eur.\ Phys.\ J.\ C {\bf 22} (2001) 493,
J.~A.~Oller and E.~Oset,
Phys.\ Rev.\ D {\bf 60} (1999) 074023.
  F.~E.~Close and N.~A.~Tornqvist,
  J.\ Phys.\ G {\bf 28}, R249 (2002).


\bibitem{Geng:2008ag}
  L.~S.~Geng, E.~Oset, J.~R.~Pelaez and L.~Roca,
  Eur.\ Phys.\ J.\  A {\bf 39}, 81 (2009)



\bibitem{Hanhart:2008mx}
  C.~Hanhart, J.~R.~Pelaez and G.~Rios,
  Phys.\ Rev.\ Lett.\  {\bf 100}, 152001 (2008)

 \bibitem{Weinberg}
 D.~Morgan,
  Nucl.\ Phys.\  A {\bf 543} (1992) 632;   D.~Morgan and M.~R.~Pennington,
  Phys.\ Rev.\  D {\bf 48} (1993) 1185.


 \bibitem{baru}
  V.~Baru et al.,
   Phys.\ Lett.\  B {\bf 586} (2004) 53.

\bibitem{vanBeveren:2002gy}
  E.~van Beveren et al.,
  AIP Conf.\ Proc.\  {\bf 660}, 353 (2003);
  Phys.\ Rev.\  D {\bf 74}, 037501 (2006).


\bibitem{FernandezFraile:2007fv}
  D.~Fernandez-Fraile, A.~Gomez Nicola and E.~T.~Herruzo,
  Phys.\ Rev.\  D {\bf 76}, 085020 (2007)


 \bibitem{lattice1}
  Ph.~Boucaud {\it et al.}  [ETM Collaboration],
  Phys.\ Lett.\  B {\bf 650}, 304 (2007)
  C.~Allton {\it et al.}  [RBC and UKQCD Collaborations],
  Phys.\ Rev.\  D {\bf 76}, 014504 (2007)
  C.~W.~Bernard {\it et al.},
  Phys.\ Rev.\  D {\bf 64}, 054506 (2001)
  C.~R.~Allton {\it et al.}
  Phys.\ Lett.\  B {\bf 628}, 125 (2005)
  M.~Gockeler {\it et al.}
  [QCDSF Collaboration],
  [arXiv:hep-lat/0810.5337].

\bibitem{bruns}
  P.~C.~Bruns and U.-G.~Mei\ss ner,
  Eur.\ Phys.\ J.\  C {\bf 40} (2005) 97.

\bibitem{mol}
S. Weinberg, Phys. Rev. {\bf
130}, 776 (1963);   Y.~Kalashnikova et al.,
  Eur.\ Phys.\ J.\  A {\bf 24} (2005) 437.

\bibitem{Guo:2008nc}
  F.~K.~Guo, C.~Hanhart, F.~J.~Llanes-Estrada and U.~G.~Meissner,
  Phys.\ Lett.\  B {\bf 678}, 90 (2009)

\bibitem{Prelovsek:2010kg}
  S.~Prelovsek, {\it et al.}
  arXiv:1005.0948 [hep-lat].
  S.~Prelovsek and D.~Mohler,
  Phys.\ Rev.\  D {\bf 79}, 014503 (2009)


\end{thebibliography}
\end{document}